# Electronic confinement in quantum dots of twisted bilayer graphene

Xiao-Feng Zhou[1], Yi-Wen Liu[1], Hong-Yi Yan[1], Zhong-Qiu Fu[1,2], Haiwen Liu[1], and Lin He[1,*]

[1]Center for Advanced Quantum Studies, Department of Physics, Beijing Normal University, Beijing, 100875, People's Republic of China

[2]College of Education for the Future, Beijing Normal University, Zhuhai, 519087, People's Republic of China

[*]Correspondence and requests for materials should be addressed to L.H. (e-mail: helin@bnu.edu.cn).

**Electronic properties of quantum dots (QDs) depend sensitively on their parent materials. Therefore, confined electronic states in graphene QDs (GQDs) of monolayer and Bernal-stacked bilayer graphene are quite different. Twisted bilayer graphene (TBG) is distinct from monolayer and Bernal-stacked bilayer graphene because of the new degree of freedom: twist angle. In the past few years, numerous efforts have been made to realize the GQDs of monolayer and Bernal-stacked bilayer graphene and achieved great success. Thus far, however, strategies for realizing GQDs of TBG have been elusive. Here, we demonstrate a general approach for fabricating stationary GQDs of TBG by introducing nanoscale *p-n* junctions with sharp boundaries in the TBG. We verify the confinement of low-energy massless Dirac fermions via whispering-gallery modes in the GQDs of TBG. Unexpectedly, electronic states around van Hove singularities of the TBG are also strongly modified around the GQDs. Such a feature has never been reported and is attributed to spatial variation of the interlayer coupling in the TBG induced by the GQDs.**

Graphene quantum dots (GQDs) have attracted extensive attention because of their novel quantum states, and the GQDs of monolayer and Bernal-stacked bilayer graphene are widely explored over the past few years [1-29]. The confined electronic states in the GQDs of monolayer and Bernal-stacked bilayer graphene exhibit quite different electronic properties due to the distinct nature of their charge carriers. By introducing a twist angle, twisted bilayer graphene (TBG) becomes a unique system with continuous tunability of structures and electronic bands [30-35]. Although the fabrication of the GQDs of monolayer [1-11,13,14,16-21] and Bernal-stacked bilayer graphene [22-29] has achieved great success and many efforts have been made to study the TBG very recently [36-38], study of GQDs of the TBG has never been reported up to now.

In this work, we demonstrate a general approach for fabricating stationary GQDs of the TBG. The TBG with different twist angles are grown on a Cu/Ni alloy through chemical vapor deposition (CVD) method and nanoscale monolayer S islands are introduced between the TBG and the substrate to generate nanoscale *p-n* junctions with sharp boundaries in the TBG. The formation of the GQDs of the TBG is verified by using scanning tunneling microscopy/spectroscopy (STM/STS) and low-energy massless Dirac Fermions of the TBG are temporarily confined in the GQDs via whispering-gallery modes (WGMs). In the TBG, the interlayer coupling between the two graphene layers will generate two low-energy van Hove singularities (VHSs). Our result, unexpectedly, reveals that the electronic states around the VHSs are strongly modified around the GQDs, which is attributed to spatial variation of the interlayer coupling in the TBG around the nanoscale *p-n* junctions.

Figure 1(a) shows a schematic of the strategy for fabricating stationary GQDs of the TBG. Firstly, Ni atoms were electroplated onto both sides of a S-rich Cu foil to obtain the Cu/Ni foil. The solubility of C in Ni is higher than that in Cu at the temperature for graphene growth (about 1050 ℃). Therefore, the mixture of Cu and Ni can adjust the amount of solubilized C in the substrate and, consequently, control the thickness of graphene [39]. In our experiment, the Ni content of the substrate is about 16%, which is demonstrated to prefer to grow large area bilayer graphene [39]

(see Fig. S1 and S2 of the Supplemental Material for the SEM and optical microscope images showing the layers of the graphene on our Cu/Ni substrate [40]). Then, the TBG was grown on the substrate via a low-pressure CVD method (see Methods of the Supplemental Material for experimental details [40]). During the growth process, the S atoms segregate from the substrate and form monolayer S islands at the interface between the TBG and the substrate [3]. The interfacial nanoscale S islands are expected to introduce sharp electronic junctions and, consequently, generate nanoscale GQDs in the continuous TBG.

Figure 1(b) shows a representative STM topographic image of a TBG (twist angle $\theta$ = 7.95°) with three interfacial S islands obtained in our experiment. The insets of Fig. 1(b) show the height profiles across the S islands and the heights of all these S atomic islands are measured about 200 pm, which is consistent with the thickness of monolayer S islands reported previously [3,10]. Figure 1(c) shows an enlarged image of the TBG, clearly exhibiting both the moiré patterns and atomic lattice of graphene. Fast Fourier transform (FFT) of the STM image further confirms the studied structure is TBG (Fig. 1(d)). The bright spots in the yellow dashed circles represent the reciprocal lattice of graphene and the bright spots in the blue dashed circles represent the reciprocal lattice of the moiré structure of the TBG. Figures 1(e)-1(h) show typical STM images of several TBG with different twist angles. The interfacial nanoscale S islands can be always observed in the obtained TBG. In our experiment, the heights of these S islands are all about 200 pm and the radius of most of the S islands is about 3-8 nm. The above result demonstrate explicitly that we successfully synthesized the TBG with interfacial nanoscale S islands according to the strategy shown in Fig. 1(a).

To confirm the formation of GQDs in the TBG, we carried out both field-emission resonances (FER) and STS measurements of the TBG on and off the nanoscale S islands. Figure 2 summarizes representative results obtained in a 7.23° TBG on and off a S island with the radius of about 4.7 nm (see Fig. S3 of the Supplemental Material for more details of this region [40]). According to the energy shifts in the FER peaks (Fig. 2(d)), *i.e.*, the peaks in d$z$/d$V$ curves, there is a large difference in the local work function, about 0.46 eV, of the TBG on and off the S island. Such a result demonstrates explicitly that the interfacial S island generates large electronic junctions in the TBG. The formation of the GQD in the TBG is further confirmed in our STS measurements, as shown in Fig. 2(c). The spectra recorded off the GQD show that the Dirac point (DP) of the TBG is about -0.2 eV ($n$ doped) and there are two low-energy VHSs at -0.75 eV and 0.28 eV, labeled as $VHS_{II}$ and $VHS_I$ respectively. In the TBG, the twisting $\theta$ leads to a relative shift of the two Dirac cones $|\Delta K| = 2|K|\sin(\theta/2)$, where $K$ is the reciprocal-lattice vector, in reciprocal space (Fig. 2(a)). The interlayer coupling leads to the emergence of two saddle points appearing at the intersections of the two Dirac cones (Fig. 2(a)) and, consequently, generates two low-energy VHSs in the density of states (DOS). The $n$ doped nature of the TBG off the GQD is also consistent with previous studies about graphene on the Cu foil [2,6,9,11]. The spectra of the TBG recorded on the GQD show that the DP is moved to about 0.25 eV ($p$ doped) and the two VHSs are shifted about 0.46 eV to positive energy [see Fig. 2(c), for example, the energy of the $VHS_I$ is shifted from 0.28 eV off the GQD to about 0.74 eV on the GQD]. Obviously, the measured energy difference on and off the GQD by the STS spectra and the FER spectra agrees quite well with each other. Our experiment demonstrates that the interfacial monolayer S atoms generate nanoscale $p$-$n$ junctions with sharp boundary in the TBG and Fig. 2(a) shows the schematic diagram of the $n$-$p$-$n$ junction of the GQD in the TBG (see Figs. S4-S6 of the Supplemental Material for more STS spectra [40]).

The most pronounced feature observed in the spectra on the GQD of the TBG is the emergence of several resonant peaks below the DP, as shown in Fig. 2(c). These peaks are attributed to the formation of quasibound states on the GQD due to the confinement of massless Dirac fermions of the TBG. Theoretical calculations predict that the low-energy quasiparticles in the TBG are massless Dirac fermions and have the same chirality as that in the monolayer graphene [35]. Therefore, the low-energy quasiparticles in the TBG should be temporarily confined in the GQDs via the WGMs due to the Klein tunneling of the massless Dirac fermions [3,4,6,8,10]. For the quasibound states confined via the WGMs, the energies should be equally spaced, as observed in our experiment (the peaks that are labeled as $N_1$-$N_3$). Simultaneously, the average level spacing of the quasibound states can be expressed as $\Delta E \approx \alpha \hbar v_F / R$ ($\Delta E$ is the average energy spacing of the quasibound states, α is a dimensionless constant of order unity, $\hbar$ is the Planck's constant divided by 2π, $v_F = 1 \times 10^6 m/s$ is the Fermi velocity, and $R$ is the radius of the quantum dot) [1,3,4,6]. Such a result is also demonstrated in the GQDs of the TBG in our experiment (Fig. S7 of the Supplemental Material [40]). The formation of quasibound states in the GQD of the TBG via the WGMs is further explicitly confirmed by carrying out spatial-resolution STS maps measurement, which reflects the local DOS (LDOS) in real space at selected energies. Figure 3 shows STS maps at energies of the $N_1$, $N_2$, and $N_3$ states. For the lowest quasibound state (the $N_1$ state), the LDOS exhibits a maximum in the center of the GQD. With increasing energy of the quasibound states, the LDOS display shell structures with the maxima progressively approaching the boundary of the GQD. These results are the characteristic feature of the quasibound states confined via the WGMs [3,4,6,8,10,20].

Besides the formation of the low-energy quasibound states, it is surprising to find that the electronic states around the VHSs of the TBG are also strongly modified by the GQDs. To clearly show this effect, we plot the deviation of the conductance map around the $VHS_I$ measured across the GQD, as shown in Fig. 4(a). In the GQD of graphene monolayer, the nanoscale *p-n* junctions are expected to mainly confine the low-energy quasiparticles below the DP and the electronic states above the DP should be weakly affected. However, this is not the case in the TBG and the electronic states around the VHSs seems strongly affected around the *p-n* junctions. To further explore the origin of the observed phenomenon, we carry out STS maps at different energies around the $VHS_I$, as shown in Figs. 4(b)-(e) (see Figs. S8, S9, S10 of the Supplemental Material [40] for results obtained around the $VHS_{II}$ and obtained in different GQDs of the TBG, exhibiting the same features as that shown in Fig. 4). Obviously, these high-energy electronic states show similar features as the low-energy quasibound states (Fig. 3). However, there is a main difference between them: the electronic states at different energies around the VHSs could exhibit distribution extending outside the GQD, unlike the low-energy quasibound states that are confined within the GQD. In the TBG, the VHSs arise from the interlayer coupling between the two graphene sheets and, therefore, they are sensitively affected by the interlayer coupling strength. Around the VHSs, the density of states become more prominent, as shown in Fig. 4 (a). To be specific, outside the GQD the remarkable DOS presents around 0.3 eV, while within the GQD the prominent DOS presents around 0.7 eV, and there exists a smooth crossover around the boundary of the GQD. The spatial DOS distributions around the VHS are shown in Figs. 4 (b)-(e). The peak of DOS gradually shifts from outside the GQD to its inner part when enlarging the energy. The spatial modulation of DOS within and outside the GQD is clearly captured by the values of $VHS_I$ (as shown in Fig. 2(a)). The observed smooth crossover of DOS around the boundary of the GQD originates from the spatially modified interlayer coupling strength around the boundary, which is due to the slightly varying interlayer distance between the two-twist graphene sheet. Around the GQD, the interfacial monolayer S island may spatially modify the interlayer coupling strength. This is especially true around the boundary of the island because the interlayer distance between the two twist graphene sheet may vary slightly around the boundary of the interfacial S island.

In summary, we report a general approach for fabricating stationary GQDs of the TBG by introducing interfacial monolayer S islands between the TBG and the supporting substrate. The interfacial S islands generate nanoscale *p-n* junctions with sharp boundaries in the TBG and help to realize the GQDs in a continuous TBG. The GQDs confine the low-energy massless Dirac fermions of the TBG into quasibound states via the WGMs. Unexpectedly, electronic states around van Hove singularities of the TBG are also strongly modified around the GQDs, which is attributed to spatial variation of the interlayer coupling in the TBG induced by the interfacial monolayer S islands.

**Acknowledgements**

This work was supported by the National Natural Science Foundation of China (Grant Nos. 11974050, 11674029, 11921005) and National Key R and D Program of

China (Grant No. 2017YFA0303301). L.H. also acknowledges support from the National Program for Support of Top-notch Young Professionals, support from "the Fundamental Research Funds for the Central Universities", and support from "Chang Jiang Scholars Program".

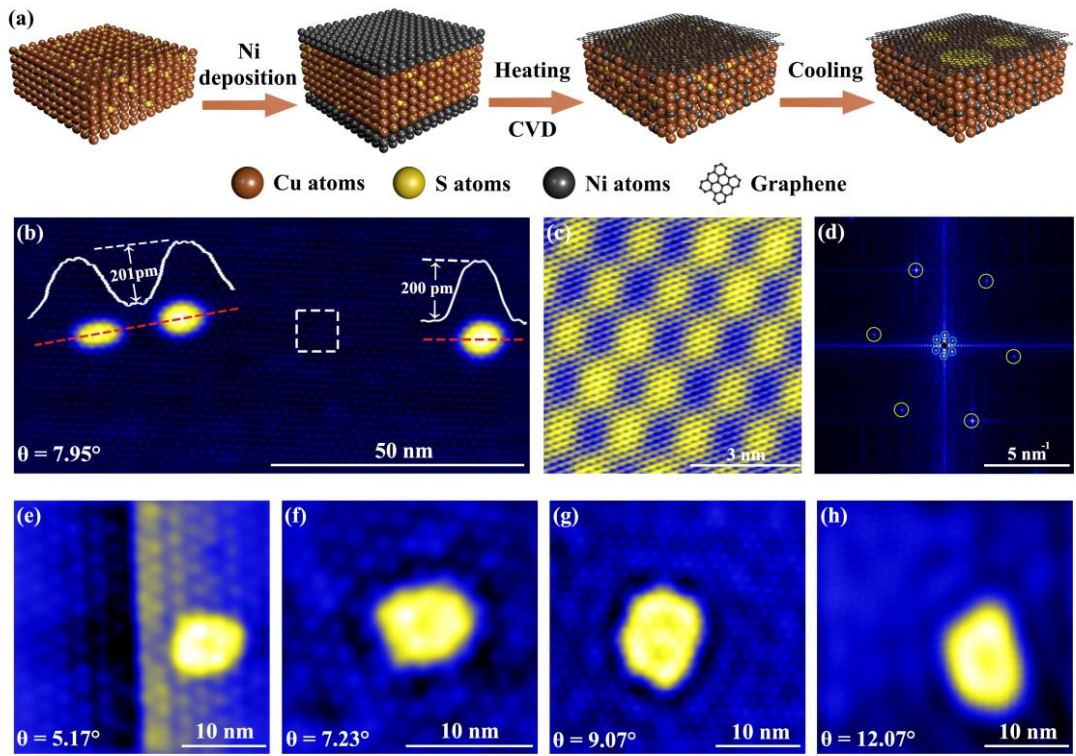

**FIG. 1. Fabrication and characterization of stationary GQDs of the TBG.** (a) The schematic of the strategy for fabricating stationary GQDs of the TBG. (b) A representative STM topographic image of a TBG region with three QDs obtained in our experiment ($V_s$ = 0.8 V, $I$ = 100 pA). Inset: the height profiles across the S islands along the red dashed lines. (c) A zoom-in atomic-resolution STM image in white dashed square in panel (b) ($V_s$ = 0.8 V, $I$ = 100 pA), which exhibits both the moiré patterns and atomic lattice of graphene. (d) FFT image of panel (c) shows both reciprocal lattice of graphene (yellow circles) and reciprocal lattice of moiré pattern (blue circles). (e-h) The typical STM topographic images of several GQDs of TBG with different twist angles. The scanning parameters are $V_s$ = 1 V, $I$ = 320 pA (e), $V_s$ =

0.5 V, $I$ = 320 pA (f), $V_s$ = 1.2 V, $I$ = 320 pA (g), $V_s$ = 0.5 V, $I$ = 320 pA (h), respectively.

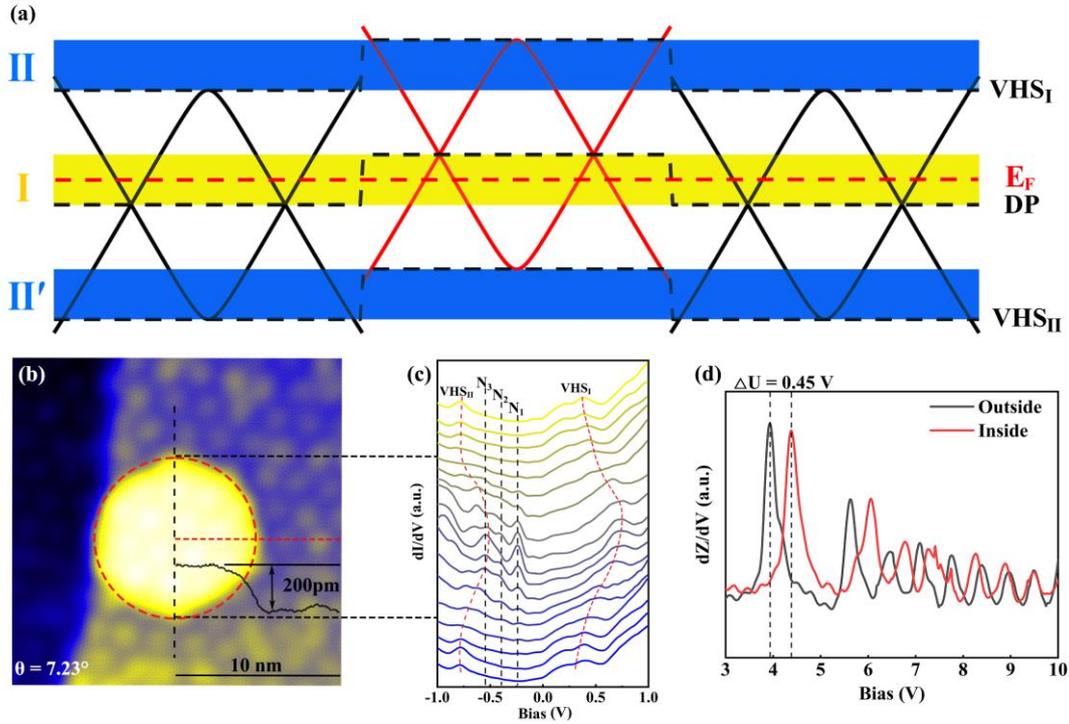

**FIG. 2. The FER and STS measurements on and off a GQD of the 7.23 ° TBG.** (a) Schematic diagram of *n-p-n* junction and electronic band structures of the TBG on and off the GQD. Off the GQD, the TBG is *n* doped, while, on the GQD, the TBG is *p* doped. The yellow region I represents the movement of the DP and the blue regions II and II' represent the movement of the $VHS_I$ and $VHS_{II}$, respectively. (b) A STM image of a GQD with the radius about 4.7 nm of the 7.23 ° TBG. The red dashed circle represents the outline of the GQD. Inset: the height profiles across the S islands along the red dashed line. (c) The STS spectra recorded across the GQD along the vertical black dashed line in (b) ($V_s$ = 1.0 V, $I$ = 100 pA). The spectra show that the DP and two VHSs of TBG are shifted about 0.45 eV. Besides, there are several resonant peaks below the DP in the spectra of the TBG on the GQD. (d) FER measured on and off the GQD of TBG. The peaks in d$Z$/d$V$ curves show a large difference in the local work function, about 0.46 eV, on and off the GQD of TBG.

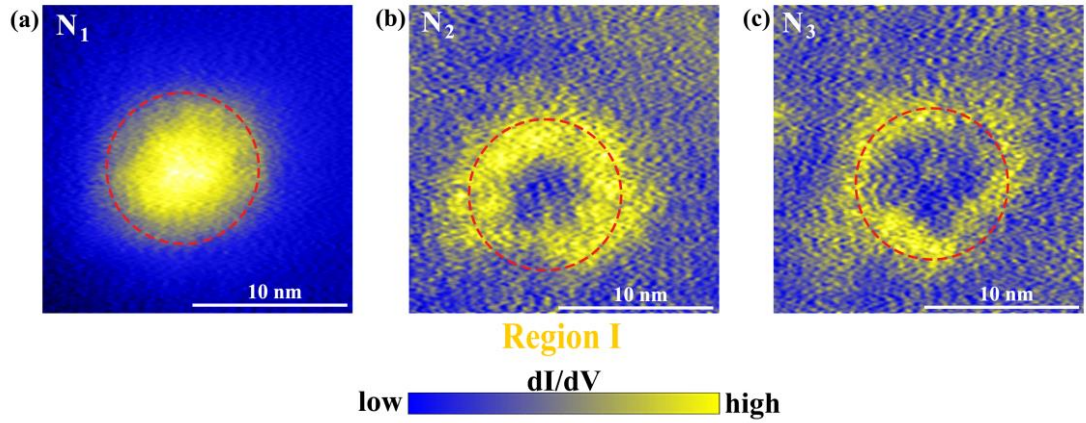

**FIG. 3. STS maps of the GQD recorded at different resonance energies.** (a) Resonance peaks $N_1$, $V_s = 0.234$ V. The maximum of the LDOS is in the center of the GQD. (b) Resonance peaks $N_2$, $V_s = 0.39$ V. (c) Resonance peaks $N_3$, $V_s = 0.538$ V. The LDOS maxima progressively approach the boundary of the GQD with increasing energy of the quasibound states, as shown in (b) and (c). The red dashed circles represent the outline of the GQD.

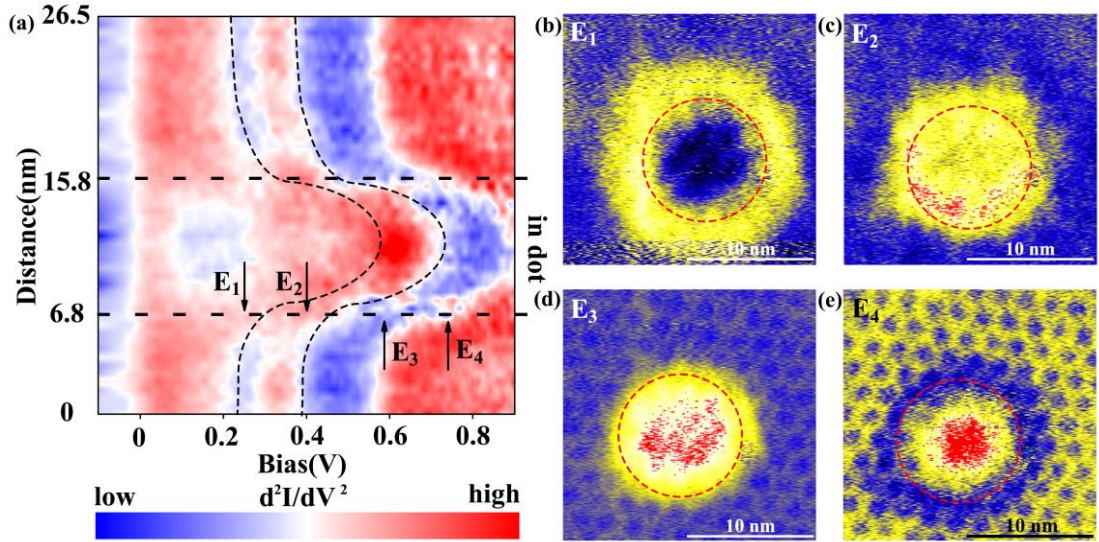

**FIG. 4. The electronic states around the VHS$_I$ of the TBG.** (a) The deviation of the conductance map measured around the VHS$_I$ across the GQD. Two horizontal black dashed lines roughly indicate the boundary position of the GQD. The modulation of the electronic states around VHS$_I$ is depicted by two black dashed curves. Four representative energies are indicated by the arrows. (b-e) The STS maps of $E_1$-$E_4$ marked in (a). Red dashed circles represent the outline of the GQD. The electronic states near the $E_1$ exhibit ring-like distribution outside the outline of the GQD, while the electronic states near the $E_4$ exhibit maximum in the center of the GQD. The $V_s$ is 0.25 V (b), 0.4 V (c), 0.59 V (d) and 0.74 V (e) respectively.